\documentclass[10pt]{article}
\begin{document}
\centerline {{\Large Conservation laws. Their role in evolutionary processes}}  
\centerline  {{\large (The method of skew-symmetric differential forms)}}
\centerline {\it L.~I. Petrova }
\centerline{{\it Moscow State University, Russia}}
\centerline{E-mail: ptr@cs.msu.su}
\bigskip

The apparatus of skew-symmetric differential forms enables us to understand   
the meaning of the conservation laws and to see their peculiarities. 
This apparatus discloses the controlling role of the conservation laws in 
evolutionary processes, which proceed in material media and lead to origination 
of various structures and forming physical fields. 

\section{Meaning of the concept ``conservation laws"}

The first formulations of the conservation laws for energy and linear
momentum date from the 17th century. In the implicit form the conservation laws
were used by G.Galilei and by the Hollandian scientist Ch.Huygens. In 1695 
the German scientist G.Leibnitz formulated the theorem about the vital force:
``The product of the force and the displacement  gives the increase of the vital force", 
this is regarded as the first formulation of the conservation law for energy
(the term ``energy" for the vital force had been applied by T.Young 112 years
later in 1807). In the modern form this theorem is formulated as follows:
``The differential of the vital force (the kinetic energy) is equal to the 
work of forces acting along the real displace" and it can be written as [1]
$$
dT=X_i dx^i\eqno(1.1)
$$
here $X_i$ are the components of exterior force, $T=mV^2/2$ is the vital
force, and $V$ is the velocity.

(It seems that there is some ambiguity here. On the one hand, this theorem may
be regarded as the first formulation of the energy conservation law because
the energy is included into it, and on the other hand, this may be regarded as
the first formulation of the linear momentum conservation law because
the force components are included into it. Such an ambiguity and the meaning
of this theorem will be explained below).

The conservation laws for nonmechanical systems were obtained in the 19th century. 
By the German physicist R.Mayer, the English physicist J.Joule, and the German scientist
H.Helmholts it was formulated the first principle of thermodynamics that is
connected with the energy conservation law. It can be written as [2,3]
$$
dE+dw=\delta Q\eqno(1.2)
$$
where $dE$ is the change of energy, $dw$ is the work executed by the system,
$dQ$ is the heat delivered to the system.

(Here the ambiguity is also observed;
this relation includes not only energy, but the mechanical work as well.
This ambiguity will be explained below).

The second principle of thermodynamics, which is also connected with the
conservation laws, was formulated by the German physicist
R.Klausius in 1850 (on the basis of the results obtained by the French engineer
S.Carnot in 1824) and the English physicist W.Tompson in 1851. In the case of 
{\it local equilibrium} the second principle of thermodynamics can be written 
as follows [2]
$$
dS=(dE+dw)/T\eqno(1.3)
$$
Here $T$ is the temperature, $S$ is the entropy.

A bit later the conservation laws that express
a conservativity of some physical quantities of objects were discovered.
Such conservation laws can be called exact ones. As an example of the
formulation of such conservation laws
it can serve the Noether theorems [4] that under some conditions can be written as
$$
d\omega=0\eqno(1.4)
$$

(It is of interest to call attention to the fact that formulas (1.1)--(1.4)
have the form of relations in differential forms.) 

Furthermore, there are other conservation laws, that are the conservation
laws for energy, linear momentum, angular momentum, and mass. These laws
establish a balance
between variations of physical quantities and appropriate external action.
Such conservation laws can be referred to as the balance conservation laws.
These conservation laws are described by differential equations [3,5]. (Below we
shall demonstrate a connection of these differential equations, which describe
the balance conservation laws, with the differential forms).

What  meaning does the concept of ``conservation law" contain?

Owing to development of science the concept of ``conservation laws"
in thermodynamics, physics and mechanics contains different meanings.

In areas of physics related to the field theory and in the theoretical
mechanics ``the conservation laws" are those according to which there exist
conservative physical quantities or objects. These are the conservation laws
that were  called above as ``exact".

In mechanics and physics of continuous media the concept of ``conservation laws"
is related to the conservation laws for energy, linear momentum, angular momentum,
and mass that establish the balance between the change of physical quantities
and external action. These are the balance conservation laws.

In theoretical mechanics the conservation laws (except the conservation
laws that may be called exact ones) are associated with the theorem of the
vital force.

In thermodynamics the conservation laws are associated with the principles
of thermodynamics.

Thus, the concept of ``the conservation laws" is connected with exact
conservation laws, balance conservation laws and some regularities
expressed in terms of the theorem of the vital force and the principles of
thermodynamics.

Now the question arises whether there is any connection between them.

The mathematical apparatus of the skew-symmetric  differential forms
allows us to answer this question [6].

Below it will be shown that the balance conservation laws are those for
material media. The exact conservation laws are the conservation
laws for physical fields
(they correspond to physical structures which constitute physical fields).

In addition it will be shown that the balance and exact conservation
laws are related to each other. The exact conservation laws are obtained 
from the balance conservation laws as a result of interactions of the balance 
conservation laws (which appear to be noncommutative) between them. 

The principles of thermodynamics and the theorem of
vital force are just examples of taking into account such interaction
(mutual influence). The principles of
thermodynamics and the theorem of vital force integrate two balance
conservation laws, namely,
the balance conservation law for energy and that for linear momentum.

Thus, we have to distinguish two types of the conservation laws that for
convenience can be called the exact conservation laws and the balance ones.

\section{Exact conservation laws}

{\it The exact conservation laws are those that state the existence of
conservative physical quantities or objects. The exact
conservation laws are related to physical fields}. $\{$The physical fields [7] 
are a special form of the substance, they are carriers of various interactions
such as electromagnetic, gravitational, wave, nuclear and other kinds of
interactions.$\}$

The closed exterior differential forms correspond to the exact 
conservation laws. (The information about the exterior differential forms 
and appropriate bibliography one can find in [8]).

If $\theta^p$ is exterior differential form of degree $p$ ($p$-form),   
the closure conditions of the exterior differential form[] (the form differential 
is equal to zero) can be written as follows:
$$
d\theta^p=0\eqno(2.1)
$$
From this relation one can see that the closed form is a conservative 
quantity. This means that such a form can correspond to the conservation law, namely, 
to some conservative physical quantity. (In relation (2.1) the exterior differential 
form is an exact one).
 
If the exterior differential form is closed on a pseudostructure only 
(i.e. it is the closed {\it inexact} differential form), the closure
condition is written as
$$
d_\pi\theta^p=0\eqno(2.2)
$$
And the pseudostructure $\pi$ obeys the condition
$$
d_\pi{}^*\theta^p=0\eqno(2.3)
$$
where ${}^*\theta^p$ is the dual form. 

From conditions (2.2) and (2.3) one can see that the exterior differential 
form closed on pseudostructure is a conservative object, namely, this
quantity conserves on pseudostructure. This can also correspond to
some conservation law, i.e. to conservative object.

The closure conditions for the exterior differential
form ($d_{\pi }\,\theta ^p\,=\,0$)
and the dual form ($d_{\pi }\,^*\theta ^p\,=\,0$) are
mathematical expressions of the exact conservation law.

The pseudostructure and the closed exterior form defined on 
the pseudostructure make up a binary differential and geometrical structure. 
Such a binary object can be named a Bi-Structure. It is evident that such 
a structure does correspond to the exact conservation law.

The physical structures, from which physical fields are formed, are preciely  
structures that correspond to the exact conservation law.

The problem of how these structures arise and how
physical fields are formed will be discussed below. 

Relations, which specifies the physical structures ($d_{\pi }\,\theta ^p\,=\,0$,
$d_{\pi }\,^*\theta ^p\,=\,0$), turn out to coincide with the mathematical
expression for the exact conservation law.

The mathematical expression for the exact conservation law and its connection
with physical fields can be schematically written in the following way 
$$
\def\\{\vphantom{d_\pi}}
\cases{d_\pi \theta^p=0\cr d_\pi {}^{*\mskip-2mu}\theta^p=0\cr}\quad
\mapsto\quad
\cases{\\\theta^p\cr \\{}^{*\mskip-2mu}\theta^p\cr}\quad\hbox{---}\quad
\hbox{physical structures}\quad\mapsto\quad\hbox{physical fields}
$$

It is evident that the exact conservation law is that for physical fields.

Since the relations for exact conservation laws and for relevant physical 
structures (that form physical fields) are expressed in terms of the closed exterior 
and dual forms, it is evident the the field theories (which describe physical 
fields) are based on the mathematical apparatus of the closed exterior differential and 
dual forms. 

One can express the field theory operators in terms of following operators 
of exterior differential forms: 
$d$ (exterior differential), $\delta$ (the operator of transforming the form 
of degree $p+1$ into the form of degree $p$), $\delta '$ (the operator of 
cotangent transforms), $\Delta $ (that of the transform $d\delta-\delta d$), 
$\Delta '$ (the operator of the transform $d\delta'-\delta' d$). 
In terms of these operators, which act onto exterior forms, one can write down 
the operators by Green, d'Alembert, Laplace and the operator of canonical 
transform [9]. 

Eigenvalues of these operators reveal themselves 
as conjugacy conditions for the differential form elements. 

The equations that are equations of the existing field theories are those 
obtained on the basis of the properties of the exterior differential form 
theory. 

It can be shown that to the quantum mechanical equations (to the equations by 
Shr\H{o}dinger, Heisengerg and Dirac) there correspond the closed exterior 
forms of zero degree or the relevant dual forms. The closed exterior form 
of zero degree corresponds to the Schr\H{o}dinger 
equation, the close dual form corresponds to the Heisenberg equation. 
It can be pointed out that, whereas
the equations by Shr\H{o}dinger and Heisenberg describe the behavior of
the potential obtained from
the zero degree closed form,  Dirac's {\it bra-} and {\it cket}- vectors
constitute a zero degree closed exterior form itself as the result of 
conjugacy (vanishing the scalar product).

The Hamilton formalism is based on the properties of closed exterior and dual 
forms of the first degree. The closed exterior differential form 
$ds=-Hdt+p_j dq_j$ (the Poincare invariant) corresponds to field equation [4]. 

The properties of closed exterior and dual forms of the second 
degree lie at the basis of the electromagnetic field equations. The Maxwell 
equations may be written as  
$d\theta^2=0$, $d^*\theta^2=0$ [11], where $\theta^2=
\frac{1}{2}F_{\mu\nu}dx^\mu dx^\nu$ (here $F_{\mu\nu}$ is the strength tensor). 

Closed exterior and dual forms of the third degree correspond to the 
gravitational field. 

From the above stated one can see that to each type of physical fields 
there corresponds a closed exterior form of appropriate degree. 

(However, to the physical field of given type it can correspond closed 
forms of less degree. They are obtained from differential forms of the 
degree corresponding to the physical field of given type.  
In particular, to the Einstein equation for gravitational field it 
corresponds the first degree closed form, although it was pointed out that 
the type of a field with the third degree closed form 
corresponds to the gravitational field.) 

The above presented connection between the field theory equations and closed 
exterior forms shows that it is possible to classify physical fields
according to the degree of exterior differential form. But within the framework 
of only exterior differential forms one cannot understand how this classification 
is explained. This can be elucidated only by application of differential forms 
of another type, which correspond to the balance conservation laws rather then 
to the inexact ones. 

The connection between field theory and  closed exterior differential forms 
supports the invariance of field theory.

And here it should underline that the field theories are based on the properties 
of closed {\it inexact} forms. This is explained by the fact that only inexact 
exterior forms  can correspond to the physical structures that form 
physical fields.  

I should be emphasize one more property of the exterior differential forms, 
which has a physical meaning and is connected with the exact conservation laws. 

Any closed form is a differential of the form of a lower
degree: the total one $\theta^p=d\theta^{p-1}$ if the form is exact,
or the interior one $\theta^p=d_\pi\theta^{p-1}$ on pseudostructure if
the form is inexact. From this it follows that the form of lower
degree can correspond to the potential, and the closed form by itself
can correspond to the potential force. This is an additional example showing
that the closed form can have a physical meaning. Here the two-fold nature
of the closed form is revealed, on the one hand, as a conservative locally
constant object, and on the other hand, as a potential force.

It was pointed out that the exact conservation law, as well as the closed 
exterior forms,  possesses the duality.
The physical structures that correspond to the exact conservation law
manifest themselves both as conservative objects and as carriers of
potential forces.

\section{Balance conservation laws}

{\it The balance conservation laws are those that establish the balance between
the variation of a physical quantity and the corresponding external action.
These are the conservation laws for the material systems (material media)}.

$\{$A material 
system is a variety of elements that have internal structure and interact 
to one another. As examples of material systems it may be thermodynamic, 
gas dynamical, cosmic systems, systems of elementary particles (pointed above) 
and others. The physical vacuum in its properties can be regarded as an analog 
of the material system that generates some physical fields. Any material media 
are such material systems. Examples of elements that constitute the material system 
are electrons, protons, neutrons, atoms, fluid particles, cosmic objects and 
others$\}$.

The balance conservation laws are the conservation laws for energy, linear
momentum, angular momentum, and mass.

In the integral form the balance conservation laws express the following [12]:
a change of a physical quantity in an elementary volume over a time interval
is counterbalanced by the flux of a certain quantity through the boundary 
surface and by the action of sources. While pasing to the differential 
expression the fluxes are changed by divergences.

The equations of the balance conservation laws are differential (or integral)
equations that describe a variation
of functions corresponding to physical quantities [3,5,12,13].
If the material system is not a dynamical one (as in the case
of a thermodynamic system), the equations of the balance conservation
laws can be written in terms of increments of physical quantities and
governing variables.

It appears that, even without knowledge of the concrete form
of these equations, with the help of the differential forms one can see
specific features of these equations that elucidate the properties of
the balance conservation laws. To do so it is necessary to study the conjugacy
(consistency) of these equations.

Equations are conjugate if they can be contracted into identical
relations for the differential, i.e. for a closed form.

Let us analyze the equations
that describe the balance conservation laws for energy and linear momentum.

We introduce two frames of reference: the first is an inertial one
(this frame of reference is not connected with the material system), and
the second is an accompanying
one (this system is connected with the manifold constructed of 
the trajectories of the material system elements). The energy equation
in the inertial frame of reference can be reduced to the form:
$$
\frac{D\psi}{Dt}=A_1 \eqno(3.1)
$$
where $D/Dt$ is the total derivative with respect to time, $\psi $ is the
functional
of the state that specifies the material system, $A_1$ is the quantity that
depends on specific features of the system and on external energy actions onto
the system. \{The action functional, entropy, wave function
can be regarded as examples of the functional $\psi $. Thus, the equation
for energy presented in terms of the action functional $S$ has a similar form:
$DS/Dt\,=\,L$, where $\psi \,=\,S$, $A_1\,=\,L$ is the Lagrange function.
In mechanics of continuous media the equation for
energy of ideal gas can be presented in the form [3]: $Ds/Dt\,=\,0$, where
$s$ is entropy. In this case $\psi \,=\,s$, $A_1\,=\,0$. It is worth noting 
that the examples presented show that the action functional and entropy 
play the same role.\}

In the accompanying frame of reference the total derivative with respect to
time is transformed into the derivative along the trajectory. Equation (3.1)
is now written in the form
$$
{{\partial \psi }\over {\partial \xi ^1}}\,=\,A_1 \eqno(3.2)
$$
here $\xi^1$ is the coordinate along the trajectory.

In a similar manner, in the
accompanying frame of reference the equation for linear momentum appears
to be reduced to the equation of the form 
$$
{{\partial \psi}\over {\partial \xi^{\nu }}}\,=\,A_{\nu },\quad \nu \,=\,2,\,...\eqno(3.3)
$$
where $\xi ^{\nu }$ are the coordinates in the direction normal to the
trajectory, $A_{\nu }$ are the quantities that depend on the specific
features of the system and external force actions.

Eqs. (3.2), (3.3) can be convoluted into the relation
$$
d\psi\,=\,A_{\mu }\,d\xi ^{\mu },\quad (\mu\,=\,1,\,\nu )\eqno(3.4)
$$
where $d\psi $ is the differential
expression $d\psi\,=\,(\partial \psi /\partial \xi ^{\mu })d\xi ^{\mu }$.

Relation (3.4) can be written as
$$
d\psi \,=\,\omega \eqno(3.5)
$$
Here $\omega \,=\,A_{\mu }\,d\xi ^{\mu }$ is the differential form of the
first degree.

Since the balance conservation laws are evolutionary ones, the relation
obtained is also an evolutionary relation.

Relation (3.5) was obtained from the equation of the balance
conservation laws for
energy and linear momentum. In this relation the form $\omega $ is that of the
first degree. If the equations of the balance conservation laws for
angular momentum be added to the equations for energy and linear momentum,
this form in the evolutionary relation will be the form of the second degree.
And in  combination with the equation of the balance conservation law
of mass this form will be the form of degree 3.

Thus, in the general case the evolutionary relation can be written as
$$
d\psi \,=\,\omega^p \eqno(3.6)
$$
where the form degree  $p$ takes the values $p\,=\,0,1,2,3$..
(The evolutionary
relation for $p\,=\,0$ is similar to that in the differential forms, and it was
obtained from the interaction of energy and time.)

In relation (3.5) the form $\psi$ is the form of zero degree. And in relation
(3.6) the form $\psi$ is the form of $(p-1)$ degree.

Let us show that {\it the evolutionary relation  obtained from the equation
of the balance conservation laws turns out to be nonidentical}.

To do so we shall analyze relation (3.5).

A relation can be identical one if this is a relation between measurable
(invariant) quantities or between observable (metric) objects, in other
words, between quantities or objects that are comparable.

In the left-hand side of evolutionary relation (3.5) there is a
differential that is a closed form. This form is an invariant
object. The right-hand side of relation (3.5) involves the differential form
$\omega$, that is not an invariant object because in real processes, as it is
shown below, this form proves to be unclosed.

For the form to be closed the differential of the form or its commutator
must be equal to zero (the elements of the form differential are equal to
the components of its commutator).

Let us consider the commutator of the
form $\omega \,=\,A_{\mu }d\xi ^{\mu }$.
The components of the commutator of such a form (as it was pointed above) can
be written as follows:
$$
K_{\alpha \beta }\,=\,\left ({{\partial A_{\beta }}\over {\partial \xi ^{\alpha }}}\,-\,
{{\partial A_{\alpha }}\over {\partial \xi ^{\beta }}}\right )\eqno(3.7)
$$
(here the term  connected with the nondifferentiability of the manifold
has not yet been taken into account).

The coefficients $A_{\mu }$ of the form $\omega $ have been obtained either
from the equation of the balance conservation law for energy or from that for
linear momentum. This means that in the first case the coefficients depend
on the energetic action and in the second case they depend on the force action.
In actual processes energetic and force actions have different nature and appear
to be inconsistent. The commutator of the form $\omega $ constructed of 
the derivatives of such coefficients is nonzero.
This means that the differential of the form $\omega $
is nonzero as well. Thus, the form $\omega$ proves to be unclosed and is not
a measurable quantity.

This means that the evolutionary relation
involves an unmeasurable term. Such a relation cannot be an identical one. 
(In the left-hand side of this relation it stands a differential  whereas in 
the right-hand side it stands an unclosed form that is not a differential.)

The nonidentity of the evolutionary relation means that equations of the
balance conservation laws turn out to be nonconjugated (thus, if from
the energy equation we obtain the  derivative of $\psi $ in the direction
along the trajectory and from the momentum equation we find the derivative
of $\psi $ in the direction normal to the trajectory and next we calculate
their mixed derivatives, then
from the condition that the commutator of the form $\omega $
is nonzero it follows that the mixed derivatives prove to be noncommutative).
One  cannot  convolute them into an identical relation and obtain a
differential.

As it will be shown below, the nonconjugacy of the balance conservation
law equations reflects the properties of the balance conservation laws that
have a governing importance for the evolutionary processes, namely, their
noncommutativity.

In section 1 of this paper relations (1.1) and (1.2), which correspond
to the theorem of the vital force and to the first principle of thermodynamics, 
were presented. In this connection an ambiguity was mentioned. On the one
side, they involve energy and thus are related to the energy conservation law.
On the other hand, they involve the force components and thus are connected with
the conservation law for linear momentum. Such an ambiguity is due to the fact
that these relations, as well as evolutionary relation (3.5), combine two
conservation laws: the balance law of energy conservation and that of linear
momentum conservation. These relations are example of evolutionary
relation (3.5). As well as  evolutionary relation (3.5), these relations are
nonidentical.

The nonidentity of the evolutionary relation does not mean that the mathematical
description of physical processes is not completely exact. The nonidentity
of the relation means that the derivatives, whose values correspond to real
values in physical processes, cannot be consistent because they are obtained
at the expense of external action and are unmeasurable quantities.

Thus, it was shown that from the equations, which describe the
balance conservation laws, the evolutionary relation in differential forms
is obtained. This evolutionary relation is a nonidentical one as it involves
an unclosed differential form. 

The differential form, which enters into the evolutionary relation, differs from 
the exterior differential form. The exterior differential form, as it is well 
known [14], is defined on the differentiable manifold or on the manifold, which 
locally admit one-to-one mapping into the Euclidean subspaces
and into other manifolds or submanifolds of the same dimension.
A specific feature of such manifold is the closure of its metric forms. The differential 
form that enters into the evolutionary relation is defined on the accompanying 
manifold, which is a varying deforming one, namely,
a manifold  with unclosed metric forms. The differential form defined on such 
manifold cannot be closed, and the corresponding evolutionary relation cannot 
be identical. This differential form was named the evolutionary one because 
it possesses the evolutionary properties. The properties of the evolutionary differential 
forms and features of the nonidentical evolutionary relation obtained from the 
balance conservation law equations allow us to elucidate the specific features 
of the balance conservation laws and their role in evolutionary processes.

\section{Noncommutativity of the balance conservation laws}

In section 3 titled ``Balance conservation laws" the conjugacy
(consistency) of the balance conservation law equations was studied.

Equations are conjugated ones if they can be convoluted into the identical 
relation for the differential, i.e. for the closed form.

As it was shown, the relation obtained from the balance
conservation law equations (the evolutionary relation) proves to be
nonidentical. From this relation one cannot obtain a differential. This 
means that the balance conservation law equations are not consistent.

The inconsistency of the balance conservation law equations indicates that
the balance conservation laws are not commutative, that is, the results
of action of the conservation laws depend on the order in which they operate.

The ``noncommutativity" of the balance conservation laws can be explained
in the following manner.

Suppose that first
the energetic and then the force perturbations act onto a local domain of the
material system (an element and its neighborhood). Let the local domain be 
in some state $A$ at the initial instant. According to the balance 
conservation law for energy, under exposure to the energetic perturbation
the local domain develops from the state $A$ into the state $B$.
Then according to the balance conservation law for momentum, under exposure
to the force perturbation it develops from the state $B$ into the state
$C$. Suppose now that the sequence of the actions is exchanged, namely, first
the force perturbation and then the energetic one act, and the system develops
first into any state $B'$ and then it proceeds into the state $C'$.
If the state $C'$ coincides with the state $C$ (this corresponds to
the local equilibrium state of the system), that is, the result
does not depend on the sequence of perturbations of different types (and
on the sequence of implementing the relevant balance conservation laws), then 
this means that the 
balance conservation laws are commutative. If the state $C'$ does not coincide
with the state $C$ (that is, the system state turns out to be not the
equilibrium one) this means that the balance conservation laws 
prove to be noncommutative.

Here it is to be noted that noncommutativity of the
balance conservation laws reflects the state of the material system.

The reason for noncommutativity of the balance conservation laws is connected
with the fact that the material system is subjected to actions of different 
nature, the nature of these actions is inconsistent with the nature of 
the material system. 

As noted above, each balance conservation law depends on the relevant action
(which the material system is subjected to). So, the conservation law for energy 
depends on the energetic action, the
conservation law for linear momentum depends on the force action, and so on.
In actual  processes such actions have different nature, and this is the cause
of noncommutativity of the balance conservation laws. The balance conservation
laws that depend on actions of different nature cannot be commutative.

In the evolutionary relation this property is revealed in the fact that the evolutionary 
form coefficients in the relation depend on different actions and hence cannot
become consistent: derivatives of these coefficients cannot constitute the
evolutionary form commutator that is equal to zero. Because of this
fact  the evolutionary form turns out to be unclosed, and the evolutionary 
relation proves to be nonidentical.

To what result does the noncommutativity of balance conservation laws lead?

It was mentioned above that the noncommutativity of the balance conservation 
laws is connected with a state of the material system. This is reflected by 
the evolutionary relation. (It should be emphasized that the evolutionary 
relation treats a state of the {\it local domain} of the material system.)

Let us consider  evolutionary relation (3.5).

In the left-hand side of the evolutionary relation there is the functional
expression $d\psi$ that determines the state of the material system, and in 
the right-hand side there is the form $\omega^p$ whose coefficients depend on
external actions and the material system characteristics.

If the evolutionary relation proves to be identical, one can
obtain the differential $d\psi $ and find the state function $\psi $, this
will indicate that the material system state is in equilibrium. But if the
evolutionary relation be nonidentical, this indicates an absence of the
differential $d\psi $ and nonequilibrium of the material system state.

Hereafter the differential $d\psi $ will be called the state differential as 
it specifies the material system state. If the state
differential be an exact closed form, this corresponds to the equilibrium
system state, whereas, if the state differential be an inexact closed form,
this will correspond to the locally equilibrium state.

The evolutionary relation gives a possibility to determine either 
availability or absence of the stade differential (the closed form).
And this allows us, firstly, to recognize whether the material system state is 
in equilibrium, in local equilibrium or not in equilibrium, 
and secondly, to determine the conditions of transition from one state 
into another (this explains the mechanism of such a transition).

It is evident that, if the balance conservation laws be commutative,
the evolutionary relation would be identical and from that it would be possible 
to get the state differential $d\psi $, this would indicate  that the material system 
is in the equilibrium state.

However, as it has been shown, in real processes the balance conservation laws
are noncommutative. The evolutionary relation is not identical and from this 
relation one cannot obtain the differential $d\psi $. This means that the system 
state is nonequilibrium.

The nonequilibrium state means that there is an internal force in the material system. 
It is evident that the internal force originates at the expense of some quantity
described by the evolutionary form commutator. (If the evolutionary form commutator
be zero, the evolutionary relation would be identical, and this would point to
the equilibrium state, i.e. the absence of internal forces.) Everything that
gives a contribution into the evolutionary form commutator leads to emergence 
of the internal force.

Here attention should be drawn to some properties of physical quantities
of the material system and specific features  of its variayion. It is precisely 
with these specific features the nonequilibrium state of material system 
is connected. 

As it was already pointed out, energy, linear momentum, angular momentum,
and mass are the physical quantities of the material system. They relate to
the same material system. Every material system has its own nature
described by a state function. The physical quantities have to be
expressed in terms of this state function. And for this reason they have to be
self-consistent.

Under any actions experienced by the material system its physical
quantities vary. Variation of the physical quantities proceeds according
to the balance conservation laws, each physical quantity being varied
in accordance to its own balance conservation law.

The physical quantities changed due to the action of the balance
conservation laws become inconsistent. This is connected with the
noncommutativity of the balance conservation laws and reflects the
following physical fact.

Each external action, as the result of which a change of physical
quantities has been produced, has the nature different from
that of the material system itself. For this reason the changed physical
quantities cannot directly become the physical quantities of the material
system itself. (The noncommutativity of the balance conservation laws does
not allow a direct transformation of the external actions into the physical
quantities of the material system). The changed physical quantities prove to
be inconsistent. As a result it arises an unmeasurable quantity that
is described by the commutator of the evolutionary form $\omega^p$
and acts as an internal force.

For the physical quantities of the material system itself become consistent,
the modified physical quantities have to come to agreement with the
properties of the material system. Such transitions, as it will be shown below,
are also governed by the balance conservation laws.

\section{Nonequilibrium of the material system}
What does the material system nonequilibrium?

The analysis of the nonidentical evolutionary relation allows to answer this question. 

If the relation is evolutionary and nonidentical, it is a selfvarying relation, that is,
a change of one object in the relation leads to a change of the other object,
and in turn a change of the latter leads to a change of the former and so on.
Since one of the objects in a nonidentical relation is an unmeasurable
quantity,  the latter object cannot be exactly compared with
the former. The process of mutual variation of the
objects involved into this relation takes place. The evolutionary
nonidentical relation proves to be a selfvarying one.

The evolutionary relation obtained from the balance conservation laws
possesses this specific feature. Such a specific feature of
the evolutionary relation explains the particularities of the material
system, namely, {\it the selfvariation of its nonequilibrium
state}.

The selfvariation mechanism of the {\it nonequilibrium}
state of the material system can be understood if to carry out the analysis of 
the selfvariation of the evolutionary relation. 

Here it should be noted that, if the metric forms of the manifold
on which the differential form is defined, be not closed, the
differential form commutator will involve a term with the commutator
of the manifold metric form (that specifies the manifold deformation)[8]. 
The evolutionary form in the evolutionary relation is defined on the
accompanying manifold that for real processes appears to be a deformable
manifold because it is formed simultaneously with a change of the
material system state and depends on the physical processes.
Such a manifold cannot be a manifold with closed metric forms.
Hence, the term containing the characteristics of the
manifold will be included into the evolutionary form commutator in
addition to the term connected with derivatives of the form
coefficients. The interaction between these terms of
different nature describes a mutual change of
the state of the material system.

Let us examine this with an example of the commutator of the form
$\omega \,=\,A_{\mu }\,d\xi ^{\mu }$ that is included into
the evolutionary relation (3.5).

If it is
possible to define the coefficients of connectedness
$\Gamma_{\alpha \beta }^{\sigma }$ (for a nondifferentiable
manifold they are not symmetrical ones), the form  commutator
can be written as
$$
K_{\alpha \beta }\,=\,\left ({{\partial A_{\beta }}\over
{\partial \xi ^{\alpha }}}\,-\, {{\partial A_{\alpha }}\over
{\partial \xi ^{\beta }}}\right )\,+\,(\Gamma _{\beta \alpha
}^{\sigma }\,-\, \Gamma _{\alpha \beta }^{\sigma })\,A_{\sigma }
$$
where $(\Gamma _{\beta \alpha }^{\sigma }\,-\,
\Gamma _{\alpha \beta }^{\sigma })$ is the metric form commutator 
(that specifies the manifold torsion). 

The emergence of the second term can only change the commutator and
cannot make it zero (because the terms of the commutator have
different nature). In the
material system the internal force will continue to act even in the absence
of external actions. The further deformation (torsion) of the manifold
will go on. This leads to a change of the metric form commutator, 
produces a change of the exterior form and its commutator and so on. 
Such a process is governed by the nonidentical evolutionary relation and, 
in turn, produces a change of the evolutionary relation.

The process of selfvariation of the commutator specifies a
change of the internal force that acts in the material system. This indicates 
a change of the material system state. But the material system state
remains nonequilibrium in this process because the internal forces
do not vanish due to the evolutionary form commutator remains to be nonzero.

At this point it should be emphasized that such selfvariation of the material
system state proceeds under the action of internal (rather than external)
forces. That will go on even in the absence of external forces. That is,
the selfvariation of the
nonequilibrium state of the material system takes place.

Here it should be noted that in a real physical process the internal forces
can be increased due to the selfvariation of the nonequilibrium state of the
material system.  This can lead to the development of instability in the 
material system [15]. 
\{For example, this was pointed out in the works by Prigogine
[16]. ``The excess entropy" in his
works is an analog of the commutator of nonintegrable form for the
thermodynamic system.  ``Production of excess entropy"  leads to the
development of instability\}.

\section{Transition of the material system into a locally equilibrium
state. (Degenerate transform)}

Whether the material system can get rid of
the internal force and transfer into the equilibrium state?

The equilibrium state corresponds the state differential that is a
closed inexact form.

The state differential is included into the nonidentical evolutionary relation.
But one cannot get the state differential from {\it nonidentical} relation.
However under degenerate transform the relation, which is identical on 
the pseudostructure,  can be obtained 
from the nonidentical evolutionary relation. From such relation 
the state differential on pseudostructure can be obtained.  
Such state differential will point to the locally equilibrium state of material 
system. 

How does the transition of the material system from the nonequilibrium state to a
locally equilibrium one proceed?

Let us consider nonidentical evolutionary relation (3.6).

As it was already mentioned, the evolutionary differential form $\omega^p$,
involved into this relation is an unclosed one for real processes. The
commutator, and hence the differential, of this form is nonzero. That is,
$$
d\omega^p\ne 0\eqno(6.1)
$$

Assume that from the evolutionary form $\omega^p$ involved into
this relation it was obtained closed-on-pseudostructure exterior
differential form (that is, a closed inexact form). This form must satisfy
the closure conditions:
$$
\left\{
\begin{array}{l}
d_\pi \omega^p=0\\
d_\pi{}^*\omega^p=0
\end{array}
\right.\eqno(6.2)
$$
In condition (6.1) the differential of the form $\omega^p$ is
nonzero whereas in condition (6.2) the differential equals zero.
It follows that the transition from condition (6.1) to 
condition (6.2), that would correspond to the transition
from the nonequilibrium state of the system to the locally equilibrium state,
is possible only as {\it the degenerate transform}, i.e. the
transform that does not conserve the differential.

{\it The degenerate transform is realized as the transition from the
accompanying noninertial coordinate system to the locally inertial that}.

To the degenerate transform it must correspond a vanishing of some functional 
expressions. 
Such functional expressions may be Jacobians, determinants, the Poisson
brackets, residues and others. A vanishing of these
functional expressions is the closure condition for a dual form. A vanishing 
of such functional expressions leads to identical relations for the exterior
differential forms written in terms of derivatives (like the Cauchy-Riemann
conditions, canonical relations, the Bianchi identities and so on).

The conditions of degenerate transform are connected with symmetries 
that can be obtained from the coefficients of evolutionary and dual forms 
and their derivatives.

Since the evolutionary relation describes the material systems and
the coefficients of the form 
$\omega^p$ depend on the material system characteristics, it is obvious that
the additional condition (the condition of degenerate transform) has to be
due to the material system properties. This may be, for
example, the availability of
any degrees of freedom in the system. (It is the availability of any degree
of freedom that can cause the redistribution between physical quantities, for 
example, between the energy and the linear momentum, in such a way that they
become simultaneously measurable quantities). The translational degrees of 
freedom, internal degrees of freedom of the system elements, and so on can be 
examples of such degrees of freedom.

The availability of the degrees of freedom in the material system indicates
that it is allowed the degenerate transform. But, for this to take 
place in reality it is necessary that the additional conditions
connected with the degrees of freedom of the material system be
realized. It is selfvariation of the nonequilibrium state of the material
system described
by the selfvarying evolutionary relation that could give rise to
realization of the additional conditions. This can appear only
spontaneously because it is caused
by internal (rather than external) reasons (the degrees of
freedom are the characteristics of
the system rather than of external actions).

On the pseudostructure $\pi$ evolutionary relation (3.6) transforms into
the relation
$$
d_\pi\psi=\omega_\pi^p\eqno(6.3)
$$
which proves to be the identical relation. Indeed, since the form
$\omega_\pi^p$ is a closed one, on the pseudostructure it turns
out to be a differential of some differential form. In other words,
this form can be written as $\omega_\pi^p=d_\pi\theta$. Relation (6.3)
is now written as
$$
d_\pi\psi=d_\pi\theta
$$
There are differentials in the left-hand and right-hand sides of
this relation. This means that the relation is an identical one.

It is from such a relation that one can find the state differential 
$d_\pi\psi$. 

The emergence of the differential $d_\pi\psi$ indicates that the material
system changes to the locally equilibrium state.
{\it But in this case the total state of the material system turns out to
be nonequilibrium.}

Under degenerate transform the evolutionary form differential vanishes only
along a certain direction (on the minipseudostructure) determined by the additional 
conditions, namely,
the conditions of degenerate transform. In other words, the equality of 
the interior differential to zero is realized. But in this case the total differential 
of the evolutionary form is nonzero. € nonzero value of the total differential
of evolutionary form means that the form remains unclosed. This shows
that the total state of the material system remains nonequilibrium.

Thus, from the properties of the nonidentical evolutionary relation and
those of the evolutionary form one can see that under realization of the 
additional 
condition (which is a condition of degenerate transform) the transition
of the material system state from nonequilibrium to locally equilibrium state
can be realized.
\section{Origination of the physical structures.
(Emergence of closed exterior forms)}

As it was pointed above, to the physical structure there correspond the closed inexact form. 

The identical relation obtained from the nonidentical evolutionary relation
under degenerate transform integrates the state differential and the closed
inexact exterior form. 
The availability of the state differential indicates
that the material system state becomes a locally equilibrium state (that is, 
the local domain of the system under consideration changes
into the equilibrium state). The availability of the exterior closed
inexact form means that the physical structure is present. This shows that
the transition of material system into the locally equilibrium state is
accompanied by the origination of physical structures. 

Since closed inexact exterior forms corresponding to physical structure
are obtained from the evolutionary relation for the material system, it follows
that physical structures are generated by the material systems. (This is
controlled by the conservation laws.)

In this manner the created physical structures are connected with the material
system, its elements, its local domains.

In the material system  origination of a physical structure reveals as a new
measurable and observable formation that
spontaneously arises in the material system. 
$\{${\it As the examples it can be fluctuations, pulsations, 
waves, vortices, creating massless particles.}$\}$.  
In the physical 
process this formation is spontaneously extracted from the local
domain of the material system and so it allows the local domain of material
system to get rid of an internal 
force and come into a locally equilibrium state.

The formation created in a local domain of the material system
(at the cost of an unmeasurable quantity that acts in the local domain
as an internal force)
and liberated from that, begins to act onto the neighboring local domain
as a force. This is a potential force, this fact is indicated by the double
meaning of the closed exterior form (on the one hand, a
conservative quantity, and, on other hand, a potential force).
The neighboring domain of the material system works over this action
that appears to be external with respect to that. If in the process the
conditions of conjugacy of the balance conservation laws turn out to be
satisfied again, the neighboring domain
will create a formation by its own, and this formation will be extracted
from this domain. (If the conjugacy conditions are not realized, the
process is finished.) In such a way the formation can move
relative to the material system. (Waves are the examples of such motions).

The formation originated is a discrete change of the inherent 
(corresponding to the nature of thematerial system) quantities of 
the material system.

A discrete change of inherent quantities and the physical structure
characteristics are mutually connected. Under the transition from one physical
structure to another the conservative quantity, namely, the exterior
closed form is changed discretely.
It is well known that there is a one-to-one connection between measurable
physical quantities and the state functional.

\section{Characteristics of physical structures and a created formation. (Value 
of the evolutionary form commutator, the properties of the material system)}

\subsection*{Characteristics of a created formation: intensity, vorticity, 
absolute and relative speeds of propagation of the formation.}

It is evident that the characteristics of the formation, as well as 
those of the created physical structure, are determined by the evolutionary 
form and its commutator and by the material system characteristics.

The following correspondence between the characteristics of the
formations emerged and characteristics of the evolutionary forms, of the 
evolutionary form commutators and of the material system is established:

1) an intensity of the formation (a potential force)
$\leftrightarrow$ {\it the  value of the first term in the
commutator of a nonintegrable form} at the instant when the formation
is created;

2) vorticity  $\leftrightarrow$ {\it the second term in the commutator
that is connected with the metric form commutator};

3) an absolute speed of propagation of the created formation (the
speed in the inertial
frame of reference) $\leftrightarrow$ {\it additional
conditions connected with degrees of freedom of the material system};

4) a speed of the formation propagation relative to the material system
$\leftrightarrow $  {\it additional
conditions connected with degrees of freedom of the material system
and the velocity of the local domain elements}.

\subsection*{Characteristics of physical structures} 

Analysis of formations originated in the material system that correspond to
physical structures allows us to clarify some properties of physical 
structures.

The observed formation and the physical structure are not identical objects.
If the wave be such a formation, the structures formed by elements of 
the front of the wave under its translation are examples of the physical 
structure. (In this case the wave element is a minipseudostructure. 
While wave translating  such element forms a line or some surface that are 
examples of the pseudostructures.) The observed formations relate to the material 
systems, whereas the physical structures relate to physical fields. In this case 
they are mutually connected.   
The characteristics of physical structure, as well as the characteristics of 
the formation, are determined by the evolutionary 
form and its commutator and by the material system characteristics. 

The originated physical structure  is a realized pseudostructure and
a corresponding inexact closed exterior form. 

The equation of the pseudostructure is obtained from the condition 
of degenerate transform and is defined by the degrees of freedom 
of the material system. The pseudostructure is mutially connected with  
an absolute speed of propagation of the created formation (see point 3 
of the previous section). 

A closed exterior form corresponding to a physical structure is a conservative
quantity, to which the state differential corresponds. The differentials of 
entropy, action, potential and others are the examples of such differentials. 
These conservative quantities describe certain charges.

The physical structure possesses two more characteristics.
This is connected with the fact that {\it inexact} closed forms correspond
to these structures. Under transition from one structure to another
the conservative quantity corresponding to the closed exterior form discretely
changes, and the pseudostructure also changes discretely.

Discrete changes of the conservative quantity and pseudostructure are 
controlled by the value of the evolutionary form commutator, which the 
commutator has at
the instant when the physical structure originates. The first term of the
evolutionary form commutator obtained from the derivatives of the evolutionary 
form coefficients controls the discrete change of the conservative
quantity. As a corresponding characteristics of the created formation it serves 
an intensity of the created formation (see point 1).  
The second term of the evolutionary form commutator obtained from the 
derivatives of the metric form coefficients of the initial manifold 
controls the pseudostructure change. Spin is the example of the second 
characteristic. As a corresponding characteristics of the created formation 
it serves a vorticity of the created formation (see point 2). Spin is a characteristic 
that determines a character of the manifold deformation before origination of
the quantum. (The spin value depends on the form degree.)

A discrete change of the conservative quantity and that of the pseudostructure
produce the quantum that is obtained while going from one structure to another.
The evolutionary form commutator formed at the instant of the structure
origination determines the characteristics of this quantum. 

\subsection*{Classification of physical structures. 
Formation of physical fields. (Parameters of the closed and dual forms)}

Closed forms that correspond to physical structures are generated by
the evolutionary relation having the parameter $p$ that defines a number of
interacting balance conservation laws, that ranges from 0 to 3. Therefore, 
the physical structures can be classified by the parameter $p$. The other 
parameter is a degree of closed forms generated by the evolutionary relation.

Under degenerate transform from the nonidentical evolutionary relation one 
obtains a relation being identical on pseudostructure. It is just a relation 
that one can integrate
and obtain the relation with the differential forms of less by one degree.

The relation obtained after integration proves to be nonidentical as well. 
 
The obtained nonidentical relation of degree $(p-1)$ can be integrated once
again if the corresponding degenerate transform is realized and the identical
relation is formed.

By sequential integrating the evolutionary relation of degree $p$ (in the case 
of realization of the corresponding degenerate transforms and forming
the identical relation), one can get closed (on the pseudostructure) exterior 
forms of degree $k$, where $k$ ranges from $p$ to $0$. 
Therefore,
physical structures can be classified by the parameter $k$ as well.

Moreover, closed exterior forms of the same degree realized in spaces of
different dimensions prove to be distinguishable because the dimension of the
pseudostructures, on which the closed forms are defined, depends on the space
dimension. As a result, the space dimension also specifies the physical
structures. This parameter determines the properties of the physical structures
rather than their type.

By introducing a classification with respect to $p$, $k$,
and a space dimension we can understand the internal connection of various
physical fields and interactions.  We can also see the correspondence between 
the degree $k$ of the
closed forms realized and the type of interactions field theory. Thus, $k=0$ corresponds to
the strong interaction, $k=1$ corresponds to the weak interaction,
$k=2$ corresponds to the electromagnetic interaction, and $k=3$ corresponds
to the gravitational interaction.

Since the physical structures are generated by numerous local domains of the
material system and at numerous instants of realizing various degrees
of freedom of the material system, it is evident that they can generate fields.
In this manner physical fields are formatted.
To obtain the physical structures that form a given physical field one has to
examine the material system  corresponding to this field and the appropriate 
evolutionary relation. In particular, to obtain the thermodynamic
structures (fluctuations, phase transitions, etc) one has to analyze the
evolutionary relation for the thermodynamic systems, 
to obtain the gas dynamic ones (waves, jumps, vortices, pulsations) 
one has to employ the evolutionary relation for gas dynamic 
systems, for the electromagnetic field one must employ
a relation ob\-tained from equations for charged particles.

Here the role of the conservation laws in
evolutionary processes should be emphasized once again.
The noncommutativity of the balance conservation laws
leads to the nonequilibrium state of the material system (subjected to external 
actions). The interaction of the noncommutative balance
conservation laws controls the process of selfvarying
the nonequilibrium state of the material system that leads to the
realization of degrees of freedom of the material system (if they are
available). The realization of
degrees of freedom allows the material system to transform into
the locally equilibrium state,
which is accompanied by emergence of physical structures. The exact
conservation laws correspond to the physical structures. Here one can see a 
connection between the balance and exact
conservation laws. The physical structures that correspond to the exact
conservation laws are produced by material system in the evolutionary
processes based on the interaction of the noncommutative balance conservation
laws.

{\it The noncommutativity of the balance conservation laws
and their controlling role in the evolutionary processes, that are
accompanied by emerging  physical structures, practically
have not been taken into account in the explicit form  anywhere}. The 
mathematical apparatus of skew-symmetric differential forms enables one to take
into account and describe these points.

Here it should be emphasized the following.

The evolutionary relation nonidentity that follows from the
conservation law noncommutativity just reflects the overdeterminacy of the 
system of the balance conservation law equations.  Actually, a number of the
balance conservation law equations is equal to a number of desired physical
quantities that specify the material system. It seems that, because a number
of equations is equal to a number
of unknown quantities, from these equations one can obtain the desired
physical quantities. But here there is some refinement. Because the physical
quantities relate to the same material system, it has to be some connection
between them. (This connection is executed by the function that specifies
the system state.) And as the physical quantities are related to each other,
then the set of the balance conservation law equations proves to be 
overdetermined one. A realization of the additional
conditions, when from the nonidentical evolutionary relation it follows
the identical relation, there corresponds to that from the overdetermined set
of equations it results the set of consistent equations from which one can find
the desired physical quantities. Since the
additional conditions can be realized only in the discrete manner, the
solutions to this system may be only quantized.

In the book by A.Pais ``The Science and the Life of Albert Einstein"
the author wrote: ``He ({\it Einstein}) hoped that the idea of the
overdeterminacy will lead to
getting the discrete solutions. He also believed that from the future theory
it will be possible to derive the partly localized solutions, which would
correspond to particles that carries the quantized electric charge".

\bigskip 
Summary.

In the work it has been shown that there are two types of the 
conservation laws.

1. The conservation laws that can be called exact ones. They point to 
an avalability of some conservative quantities or objects. Such objects 
are the physical structures,  which the physical fields and relevant 
manifolds are constructed of. These are conservation laws for physical fields.

2. The conservation laws of energy, linear and angular momentum, and mass. 
These laws are conservation laws for  material systems (material media). They 
establish a balance between changes of physical quantities and external actions. 
Such conservation laws can be called as balance ones.

It has been shown that the exact and balance conservation laws execute a relation 
between the physical structures, which form physical fields, and material systems.
The physical structures, to which the exact conservation laws correspond, are 
generated by material systems in the evolutionary processes, whose moving 
force is the noncommutativity of the balance conservation laws.

These results are obtained with the help of the mathematical apparatus 
of skew-symmetric differential forms.

1. Chetaev N.~G., Theoretical Mechanics. --Moscow, Nauka, 1987 (in Russian). 

2. Haywood R.~W., Equilibrium Thermodynamics. Wiley Inc. 1980. 

3. Clark J.~F., Machesney ~M., The Dynamics of Real Gases. Butterworths, 
London, 1964. 
 
4. Encyclopedia of Mathematics. -Moscow, Sov.~Encyc., 1979 (in Russian).

5. Fock V.~A., Theory of space, time, and gravitation. -Moscow, 
Tech.~Theor.~Lit., 1955 (in Russian).

6. Petrova L.~I., Origination of physical structures. //Izvestia RAN, Fizika, 
N 1, 2003. 

7. Encyclopedic dictionary of the physical sciences. -Moscow, Sov.~Encyc., 
1984 (in Russian).

8. Petrova L.~I., Exterior and evolutionary skew-symmetric differential forms 
and their role in mathematical physics. http://arXiv.log, 
User-ID: math-ph/0310050

9. Wheeler J.~A., Neutrino, Gravitation and Geometry. Bologna, 1960.

10. Dirac P.~A.~M., The Principles of Quantum Mechanics. Clarendon Press, 
Oxford, UK, 1958.

11. Weinberg S., Gravitation and Cosmology. Principles and applications of 
the general theory of relativity. Wiley \& Sons, Inc., N-Y, 1972.

12. Dafermos C.~M. In "Nonlinear waves". Cornell University Press, 
Ithaca-London, 1974.

13. Tolman R.~C., Relativity, Thermodynamics, and Cosmology. Clarendon Press, 
Oxford,  UK, 1969.

14. Schutz B.~F., Geometrical Methods of Mathematical Physics. Cambrige 
University Press, Cambrige, 1982.

15. Petrova L.~I., On the problem of development of the flow instability. 
//Basic Problems of Physics of Shock Waves. Chernogolovka, 1987, V.~{\bf 1}, 
P.~1, 304-306 (in Russian). 

16. Prigogine I., Introduction to Thermodynamics of Irreversible 
Processes. --C.Thomas, Springfild, 1955.

\end{document}